\documentclass[aps,prd,preprint,groupedaddress]{revtex4}

\usepackage{graphicx}
\usepackage{color}

\newcommand{\eV}{\mathinner{\mathrm{eV}}}
\newcommand{\keV}{\mathinner{\mathrm{keV}}}

\newcommand{\GeV}{\mathinner{\mathrm{GeV}}}

\def\l{\left}
\def\r{\right}

\newcommand{\beq}{\begin{equation}}
\newcommand{\eeq}{\end{equation}}
\newcommand{\bea}{\begin{eqnarray}}
\newcommand{\eea}{\end{eqnarray}}

\begin{document}

\preprint{KIAS-O14002}

\title{The 3.5 keV X-ray line signature from annihilating and decaying dark matter in Weinberg model}



\author{Seungwon Baek}
\email[]{swbaek@kias.re.kr}
\affiliation{School of Physics, KIAS, Seoul 130-722, Korea}

\author{P. Ko}
\email[]{pko@kias.re.kr}
\affiliation{School of Physics, KIAS, Seoul 130-722, Korea}


\author{Wan-Il Park}
\email[]{wipark@kias.re.kr}
\affiliation{School of Physics, KIAS, Seoul 130-722, Korea}


\date{\today}

\begin{abstract}
Recently two groups independently observed unidentified  X-ray line signal at the energy
3.55 keV from the galaxy clusters and Andromeda galaxy. We show that this anomalous signal 
can be explained in annihilating dark matter model,
for example, fermionic dark matter model in hidden sector with global $U(1)_X$  symmetry proposed by Weinberg. 
There are two scenarios for the production of the annihilating dark matters.
In the first scenario the dark matters  with mass 3.55 keV decouple from the interaction with Goldstone bosons 
and go out of thermal equilibrium at high temperature ($>$ 1 TeV) when they are still relativistic, 
their  number density per comoving volume being essentially fixed to be the current value. 
The correct relic abundance of this warm dark matter is obtained by assuming that about ${\cal O}(10^3)$ relativistic degrees
of freedom were present at the decoupling temperature or alternatively large entropy production occurred at high
temperature.
In the other scenario, the dark matters were absent at high temperature, and as the universe cools down, 
the SM particles  annihilate or decay to produce the dark matters non-thermally as in `freeze-in' scenario.
It turns out that the DM production from Higgs decay is the dominant one.
In the model we considered, only the first scenario can explain both X-ray signal and relic abundance.
The  X-ray signal arises through $p$-wave annihilation of 
dark matter pair into two photons through the scalar resonance without  
violating the constraints from big bang nucleosynthesis, cosmic microwave background,
and astrophysical objects such as red giants or white dwarfs.
We also discuss the possibility that the signal may result from a decaying dark matter in a simple
extension of Weinberg model. 
\end{abstract}

\pacs{}

\maketitle

\section{Introduction}
Recently unidentified X-ray line at 3.55 keV energy has been discovered independently
by two groups~\cite{Xray_exp}. Its statistical significance over the background is 4-5$\sigma$.
One possible explanation for the anomaly may be long-searched-for dark matter (DM).
There are already many works trying to explain the line signal in dark matter models~\cite{Xray_th}.

In this paper, we show that the X-ray line signal can be obtained by DM annihilation.
Specifically we work in the annihilating dark matter model proposed by 
Weinberg~\cite{Weinberg:2013kea}. The model has global $U(1)_X$ symmetry
under which all the standard model (SM) particles are neutral and a fermionic dark matter and 
a dark scalar are charged. The dark scalar mediates
the interaction between the dark sector and the SM sector~\cite{Higgs_portal}. 
Weinberg showed that the Goldstone boson resulting from the spontaneous symmetry breaking of 
the global symmetry can mimic dark radiation and contribute to the effective neutrino number $N_{\rm eff}$. 
If the Goldstone bosons go out of thermal equilibrium at the temperature
just above the muon (electron) mass and we can get $\Delta N_{\rm eff}=0.39 (0.57)$~\cite{Weinberg:2013kea}. 
The collider signature of the Goldstone bosons and the indirect signal of DM annihilating into two photons
in light of the Fermi-LAT 130 GeV gamma-line have been studied in the model~\cite{Weinberg_app}.

If the X-ray line signal is interpreted as a DM annihilation, the DM mass should be about 3.55 keV
and the required annihilation cross section into two photons times relative velocity of the
dark matters should be~\cite{Frandsen:2014}
\begin{equation}
\sigma v_{\rm rel} =(7.1 \times 10^{-8} - 1.3 \times 10^{-6}) \; {\rm pb}.
\label{eq:X_ray}
\end{equation}
In principle the DM relic abundance can be obtained by two independent mechanisms.
In the first mechanism dark matters go out of thermal equilibrium at high temperature when they
are relativistic. This is the same with the mechanism by which the relic neutrinos are obtained.
Since the dark matter for X-ray line signal is much heavier than neutrinos,  
we need large number of relativistic degrees of freedom 
($g_*(T_f) \sim$ 3500\footnote{This number is reduced if there had been
entropy production processes such as out-of-equilibrium decay of heavy particles.}) 
at decoupling temperature $T_f$ to get the correct relic density. 
The large $g_*(T_f)$ can be obtained, for example, by assuming
new physics such as large extra dimensional model where towers of Kaluza-Klein excited states
of the SM are predicted.
In the second mechanism, even if there were no primordial dark matter relic, the dark matters
can be produced by non-thermal annihilation of the bottom quark pair or decay of Higgs boson at low temperature  
as in the freeze-in mechanism~\cite{Hall:2009}.
Although there may be constraints on annihilating dark matters, for example, 
from big bang nucleosynthesis (BBN), cosmic microwave background (CMB),
and astrophysical objects such as red giants or white dwarfs, they can be evaded in our model.

\section{The model}
The model introduces a fermionic dark matter $\psi$ and a scalar field $\varphi$  charged under a new global symmetry $U(1)_X$
with the quantum number $1$ and $2$ respectively. Then the Lagrangian invariant under $U(1)_X$ can be
written as~\cite{Weinberg:2013kea}
\bea
{\cal L} &=& {\cal L}_{\rm SM}+\partial_\mu \varphi^* \partial^\mu \varphi -\mu^2 \varphi^* \varphi 
-\lambda_\varphi (\varphi^* \varphi)^2 -\lambda_{H\varphi} H^\dagger H \varphi^* \varphi \nonumber\\
&& +\overline{\psi} i \gamma^\mu \partial_\mu \psi -m_\psi \overline{\psi} \psi 
-({f \over 2} \overline{\psi} \psi^c \varphi + h.c.),
\eea
where $H$ is the SM Higgs doublet field and the coupling constant $f$ can be taken to be positive
without loss of generality by appropriate field redefinition. To make scalar potential stable 
the quartic couplings should satisfy
\bea
|\lambda_{H\varphi}| < 2 \sqrt{\lambda_H \lambda_\varphi},
\label{eq:vacuum_stability}
\eea
where $\lambda_H$ is the quartic coupling of $H$.
After  $H$ and $\varphi$ obtain vacuum expectation values $v_H$ and $v_\varphi$, respectively, we can
expand the scalar fields as
\bea
H={1 \over \sqrt{2}}\left(\begin{array}{c} 0 \\ v_H +h
\end{array}
\right), \quad
 \varphi ={1 \over 2} (v_\varphi + \phi) e^{i \alpha/v_\varphi}, 
\eea
where $\alpha$ is the Goldstone boson. 
Then the global $U(1)_X$ symmetry is broken down to $Z_2$ symmetry under which
$\psi \to -\psi$ and all the other fields are unchanged.
As a result the $\psi$ field splits into two Majorana mass eigenstates $\psi_\pm$:
\bea
  \psi &=& {1 \over \sqrt{2}} (\psi_+ + i\psi_-), \nonumber\\
  \psi^c &=& {1 \over \sqrt{2}} (\psi_+ -i \psi_-), 
\eea
with the masses $m_{\psi_\pm} = m_\psi \pm f v_\varphi/\sqrt{2}$.
The residual $Z_2$ symmetry prevents the DM candidate for which we take $\psi_-$ from decaying. 
The real parts of the scalar fields also mix with each other by mixing angle $\alpha_H$, 
\bea
\left(\begin{array}{c} h \\ \phi \end{array} \right)=
\left(\begin{array}{cc} c_H  & s_H \\ -s_H & c_H \end{array} \right)
\left(\begin{array}{c} H_1 \\ H_2 \end{array} \right)
\equiv O \left(\begin{array}{c} H_1 \\ H_2 \end{array} \right),
\eea
where $c_H  (s_H) \equiv \cos \alpha_H (\sin\alpha_H)$ and $H_i (i=1,2)$ are the scalar mass
eigenstates with masses $m_i$. 
With the above definition, $\alpha_H<0$ for $m_2<m_1$.
We take $H_1$ ($m_1 =125$ GeV) as the Higgs particle discovered at the LHC.

\section{Annihilating DM Scenario}
\begin{figure}
\center
\includegraphics[width=8cm]{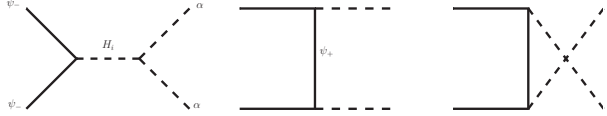}
\caption{Dominant Feynman diagrams contributing to the relic density calculation.}
\label{fig:Feyn}
\end{figure}

If the dark matters go out of thermal equilibrium in the early universe when they are still relativistic,
the current relic abundance is given by~\cite{the_early_univ}
\bea
\Omega_{\psi_-} h^2&=& 278 \frac{g_{\rm eff}}{\gamma g_{*S}(T_f)} \left(m_{\psi_-} \over {3.55 {\rm keV}}\right), \nonumber\\
 {\rm or} \quad 
Y(T_f) &\equiv& \frac{n(T_f)}{s(T_f)} =0.278 \frac{g_{\rm eff}}{g_{*S}(T_f)},
\label{eq:omega}
\eea
where $g_{\rm eff}=g {\rm (bosons)}, 3g/4 {\rm (fermion)}$ and $\gamma$ is the factor by which the entropy
per comoving volume increases by some out-of-equilibrium processes after the decoupling of $\psi_-$.
In this case the relic density is not sensitive to the decoupling temperature $T_f$, and we take
$ 10^3\, {\rm GeV}  \lesssim T_f  \lesssim 10^{10} \, {\rm GeV}$.
Taking $\Omega_{\psi_-} h^2=0.1199$, $m_{\psi_-}=3.55$ keV and $g_{\rm eff}=3/4*2=1.5$, we get
$g_{*S}(x_f)=3.4 \times 10^3/\gamma$. We can either assume large $\gamma$ or some beyond-the-standard-model
physics, for example large extra dimensional models with Kaluza-Klein towers of the SM particles with masses less
than $T_f$, to explain this rather large $g_{*S}(x_f)$.
At the freeze-out temperature $T_f$, the DM  $\psi_-$'s are decoupled from the annihilation process $\psi_- \psi_-
\to \alpha\alpha$ and go out of thermal equilibrium. This process is dominated by the Feynman diagrams shown in Fig.~\ref{fig:Feyn}.
The interactions between the Goldstone boson and dark matter/Higgs  particles are proportional
to $1/v_\varphi$. Other relevant parameters for the diagrams are the Yukawa coupling $f$,
the scalar masses $m_i (i=1,2)$. For example, the $s$-channel diagram gives the annihilation
cross section
\bea
\sigma v_{\rm rel} \approx \frac{f^2 v_{\rm rel}^2 s^2}{256 \pi v_\varphi^2 (s-m_\phi^2)^2},
\eea
where we neglected the small $\alpha_H$.
Since $\psi_-$ decouples at high temperature, we can approximate $s\approx T^2 \gg m_{\psi_-}^2 (\sim m_\phi^2)$.
Then, since the number density for relativistic particles $n\approx T^3$ and $v_{\rm rel}\approx 1$, when
\bea
 n \langle \sigma v_{\rm rel} \rangle \approx \frac{f^2 T^3}{256 \pi v_\varphi^2 }
\eea
becomes less than  the Hubble expansion parameter $H\approx T^2/m_{\rm pl}$, 
the DM particles decouple from the thermal bath. The freeze-out temperature is determined from
the condition
\bea
\frac{f^2 m_{\rm pl} T_f}{256 \pi v_\varphi^2} \approx 1.
\label{eq:psi_decoup}
\eea

\begin{figure}
\center
\includegraphics[width=6cm]{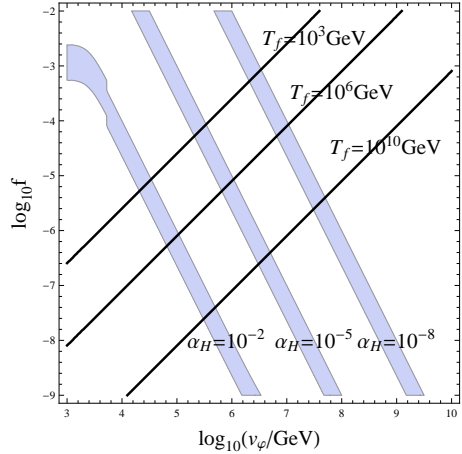}
\caption{Allowed parameter space in $(v_\varphi, f)$-plane for relic abundance and X-ray line signal. 
The colored regions can explain the X-ray line signal by $\psi_- \psi_- \to \gamma\gamma$
for the mixing angles $\alpha_H=-10^{-2},-10^{-5},-10^{-8}$ from left.
The black lines give the correct relic density $\Omega_{\psi_-} h^2=0.1199$ 
by decoupling at $T_f=10^3, 10^6, 10^{10}$ GeV from above.
}
\label{fig:VM2}
\end{figure}
Fig.~\ref{fig:VM2} shows allowed parameter space in $(v_\varphi, f)$-plane for relic abundance and X-ray line signal. 
The colored regions can explain the X-ray line signal by $\psi_- \psi_- \to \gamma\gamma$ (See (\ref{eq:XXrr}))
for the mixing angles $\alpha_H=-10^{-2},-10^{-5},-10^{-8}$ from left. We assumed the resonance condition,
$m_2 = 2 m_{\psi_-} =7.1$ keV, is satisfied.
The black lines give the correct relic density $\Omega_{\psi_-} h^2=0.1199$ by satisfying (\ref{eq:psi_decoup})
for $T_f=10^3, 10^6, 10^{10}$ GeV from above.
Large mixing angle is disfavored because the invisible Higgs decay width becomes too large, and
we restrict $\alpha_H \lesssim 0.01$.
If $\psi_+$ mass becomes similar to the mass
of the DM $\psi_-$, we can get additional coannihilation channel $\psi_- \psi_+\to \alpha \alpha$. 
However, we do not consider this case because it does not change the decoupling temperature much.

The  cross section of the DM annihilation into the SM particles always involves the mixing matrix, $O$, in the form
\bea
\left|\sum_{i=1,2} { O_{1i} O_{2i} \over s-m_i^2 +i m_i \Gamma_i} \right|^2,
\eea
where $\Gamma_i$ is the total decay width of $H_i$. 
The cross section is vanishing for $s \gg m_i^2$ due to the orthogonality of matrix $O$. 
The $s$-channel annihilation cross section $\psi_- \psi_- \to f \bar{f}$ is obtained to be
\bea
\sigma v_{\rm rel} =\frac{N_c^f f^2 m_f^2 v_{\rm rel}^2 s}{64 \pi v_H^2} 
\left|\sum_{i=1,2} { O_{1i} O_{2i} \over s-m_i^2+i m_i \Gamma_i} \right|^2
\left(1-{4 m_f^2 \over s}\right)^{3/2}.
\eea
For the $\psi_- \psi_- \to W^+,W^- (Z,Z)$ it is given by
\bea
\sigma v_{\rm rel} &=&\frac{f^2 v_{\rm rel}^2 }{32 \pi v_H^2} 
\left|\sum_{i=1,2} { O_{1i} O_{2i} \over s-m_i^2+i m_i \Gamma_i} \right|^2 \nonumber\\
&\times&\left\{m_W^4 \left(2+\frac{(s-2m_W^2)^2}{4m_W^4}\right) \sqrt{1-{4 m_W^2 \over s}}\right. \nonumber\\
&&\left.+{m_Z^4 \over 2} \left(2+\frac{(s-2m_Z^2)^2}{4m_Z^4}\right) \sqrt{1-{4 m_Z^2 \over s}}\right\}.
\eea
These forms suggest the interactions become stronger as $s$ becomes smaller due to imperfect cancellation
and $1/s$ behavior.
The $t$-channel $\psi_-$-exchanging $\psi_- \psi_- \to H_1 H_1$  annihilation cross section
\bea
 \sigma v_{\rm rel} \approx \frac{f^4 \alpha_H^4 v_{\rm rel}^4 m_{\rm pl}}{3072 \pi s}
\eea
does not cancel but is suppressed by $(f \alpha_H)^2$ compared with the $s$-channel cross section and
is always negligible.

As the universe cools down, the decoupled dark matters may annihilate into the SM particles or additional dark matter
particles may produced by pair annihilation of the SM particles, when the interactions of dark matters with
the SM particles are strong enough. Now let us consider these possibilities. The Boltzmann equation for the
number density of $\psi_-$, $n_{\psi_-}$, is written as
\bea
 \frac{d n_{\psi_-}}{d t} + 3 H n_{\psi_-} = - \langle \sigma(\psi_- \psi_- \to X \bar{X}) v_{\rm rel} \rangle
\Big[ n_{\psi_-}^2 - (n_{\psi_-}^{\rm EQ})^2 \Big],
\label{eq:Boltzmann}
\eea
where $X(\bar{X})$ represents any SM particle and $n_{\psi_-}^{\rm EQ}$ is the number density if
it were in equilibrium with the thermal bath at time $t$ (or temperature $T$). 
The temperature of $\psi_-$ is smaller than the thermal
plasma temperature, $T$, by factor $(g_{*S}(T)/g_{*S}(T_f))^{1/3}$~\cite{the_early_univ}. Therefore, we always get
$n_{\psi_-} < n_{\psi_-}^{\rm EQ}$ after the decoupling of $\psi_-$, implying that additional $\psi_-$ may be
created from thermal plasma at lower temperature but not the other way around. 
It turns out that the production yield, $Y=n_{\psi_-}/s$,
is the largest for $b \bar{b} \to \psi_- \psi_-$. The Boltzmann equation for this case
\bea
{dY \over dT} =\sqrt{\pi g_* \over 45} \frac{3 f^2 \alpha_H^2 m_b^4 m_{\rm pl}}{64 \pi v_H^2 m_{\psi_-}^4 }
\frac{K_1^2(m_b/T)}{K_2^2(m_{\psi_-}/T)}(Y^2-Y_{\rm eq}^2),
\eea
gives the correct relic abundance when $f \alpha_H \sim 10^{-6}$ assuming there is no dark matter 
at high energy.

There is another ``freeze-in'' process for the production of $\psi_-$ DM: the decay of $H_1$,
$H_1 \to \psi_- \psi_-$~\cite{Hall:2009}.
In this case the Boltzmann equation
\bea
 {dY \over dT} \simeq -\frac{m_1^2 \Gamma_{H_1 \to \psi_- \psi_-}}{2 \pi^2 s H} K_1\left(m_1 \over T\right),
\eea
with
\bea
\Gamma_{H_1 \to \psi_- \psi_-} \simeq \frac{f^2 \alpha_H^2 m_1}{32 \pi} \left(1-{4 m_{\psi_-}^2 \over m_1^2}\right)^{3/2},
\eea
gives the correct relic abundance when $f \alpha_H \sim 10^{-8}$.
Therefore we can see the Higgs decay process is more effective in producing the dark matter particles
via freeze-in mechanism.  However, as can be seen in Fig.~\ref{fig:VM2}, in most parameter space 
$f \alpha_H \sim 10^{-8}$ is large enough to allow the DM to be in thermal equilibrium with the 
Goldstone bosons at high temperature, $T>10^3$ GeV,
making this freeze-in mechanism irrelevant\footnote{In small region in Fig.~\ref{fig:VM2} for $\alpha_H=10^{-2}$,
$f \alpha_H \gtrsim 10^{-8}$ where both freeze-in and freeze-out mechanism can contribute in producing DM.}.

Now let us consider decoupling of the Goldstone bosons from the thermal bath.
The decoupling temperature was estimated by Weinberg~\cite{Weinberg:2013kea}.  
He proposed an interesting possibility that the Goldstone boson ``can be masquerading as fractional
cosmic neutrinos'' and contribute to the effective neutrino number by amount $\Delta N_{\rm eff} =0.39 (0.57)$
when they decouple just before muon (electron) pair annihilation.
Using $\lambda_{H\varphi} \approx -m_1^2 \alpha_H/v_H v_\varphi$ for small $\alpha_H$ and small $m_2$,
the condition for the Goldstones go out of thermal equilibrium just before $f-\bar{f}$ pair
annihilation becomes~\cite{Weinberg:2013kea}
\bea
\frac{\alpha_H^4 m_f^7 m_{\rm pl}}{v_H^2 v_\varphi^2 m_2^4} \approx 1.
\eea
For example, for $(\alpha_H, v_\varphi) \approx (-10^{-4}, 10^6 {\rm GeV})$ 
($(\alpha_H, v_\varphi) \approx (-10^{-2}, 10^2 {\rm GeV})$) the Goldstone bosons decouple
just before muon (electron) pair annihilation.
As $\alpha_H$ becomes smaller, the Goldstones decouple much earlier and $\Delta N_{\rm eff} \ll 1$.

\begin{figure}
\center
\includegraphics[width=8cm]{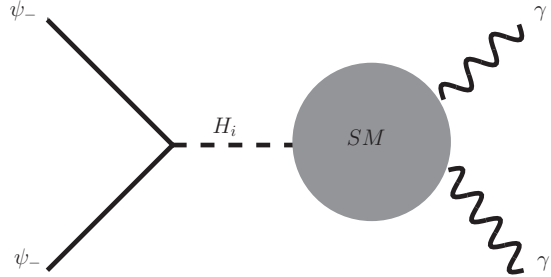}
\caption{Feynman diagram contributing to the 3.55 keV X-ray line signal.}
\label{fig:XXrr}
\end{figure}
Now let us turn to the $\psi_- \psi_- \to \gamma\gamma$ process for the 3.55 keV X-ray line signal.
The main contribution to this process is the $H_i$ mediated $s$-channel process shown in Fig.~\ref{fig:XXrr},
where a pair of
$\psi_-$'s annihilate into the dark scalar $\phi$ which mixes into the SM scalar $h$, followed by
$h$ decaying into two photons via one-loop diagram inside which $W^\pm$ and heavy quarks are
running. As a result, contrary to the $\psi_- \psi_- \to \alpha\alpha$ case for the relic density, 
$\psi_- \psi_- \to \gamma\gamma$ is very sensitive to
the mixing angle $\alpha_H$. 
The annihilation cross section for $\psi_- \psi_- \to \gamma\gamma$ is obtained to be
\bea
\sigma v_{\rm rel} &=& \frac{\alpha_{\rm em}^2 f^2 v_{\rm rel}^2 s^2}{2048 \pi^3 v_H^2}
\Bigg| \sum_{i=1,2} \frac{O_{1i} O_{2i}}{s-m_i^2 + i m_i \Gamma_i } \Bigg|^2 \nonumber\\
&\times& \Bigg|F_1\left(4 M_W^2 \over s\right) +  \sum_f N_c^f Q_f^2 F_{1/2}\left(4 m_f^2 \over s\right)\Bigg|^2,
\label{eq:XXrr}
\eea
where $v_{\rm rel}\sim 10^{-3}$ is relative velocity of dark matter and $s \approx 4 m_{\psi_-}^2/(1-v_{\rm rel}^2/4)$. 
The loop functions $F_1, F_{1/2}$ can be found in~\cite{HHG}.
To get the plot in Fig.~\ref{fig:VM2} we averaged over the velocity distribution of dark matter,
$\langle\sigma v_{\rm rel} \rangle_{v}=\int d^3 v e^{-v^2/v_0^2} \sigma v/\int d^3 v e^{-v^2/v_0^2}$ with
$v_0 \approx 10^{-3}$.
Too large mixing angle may enhance the Higgs invisible decay rates beyond the current experimental limit
through, $H_1 \to H_2 H_2 $ or $H_1 \to \psi_- \psi_-$. 
Currently at most 19\% deviation from the SM Higgs decay width ($\Gamma_{\rm SM} \approx 4$ MeV) is
allowed from the LHC Higgs search experiments~\cite{Giardino:2013bma},  and we
restrict $|\alpha_H| \leq 0.01$ in Fig.~\ref{fig:VM2}.


\subsection{Constraints}
In this section we discuss some potential cosmological and astrophysical constraints on the model.
Although the dark matter $\psi_-$ itself contributes to the effective number of neutrinos $N_{\rm eff}$~\cite{Boehm:2013jpa},
since $\psi_-$ decouples from the thermal bath at very high energy, its contribution is negligible in our model.
Another constraint may come from the injection of energy into the thermal plasma from $\psi_- \psi_- \to \gamma\gamma$ at
CMB epoch.
For $s$-wave annihilation into two photons, the CMB observation constrains
\bea
\langle \sigma v_{\rm rel} \rangle_{\rm CMB} < 2.86 \times 10^{-7} \, {\rm pb},
\eea
for a DM with mass 3.55 keV~\cite{Frandsen:2014}. In our case, the annihilation process (\ref{eq:XXrr})
is $p$-wave, and can easily avoid the above constraint because the relative velocity of DMs at CMB era was
much smaller than that at current universe.

In some astrophysical objects such as red giants, white dwarfs, and neutron stars, the plasmons may annihilate 
into a pair of DM particles through the inverse process of X-ray signal, {\it i.e.} $\gamma\gamma \to \psi_- \psi_-$.
If the produced DMs escape the objects, they could cool the stars faster than observation. This gives
the constraint on the energy loss per unit volume per unit time~\cite{Davidson:1991}
\bea
\frac{d^2 E}{d V d t} < 1.6 \times 10^6\, {\rm erg/cm^3/sec}.
\label{eq:RG}
\eea
In our model taking $\sigma v_{\rm rel} (\gamma\gamma \to \psi_- \psi_-)$ to be the same with (\ref{eq:X_ray}) for
rough estimation, we get
\bea
\frac{d^2 E}{d V d t}  &=& {1 \over 2} m_\gamma n_\gamma^2 
 \langle \sigma v_{\rm rel} (\gamma\gamma \to \psi_- \psi_-) \rangle \nonumber\\
&\sim& 10^9 \, {\rm erg/cm^3/sec},
\eea
where we have taken the parameters of red giants from \cite{Davidson:1991}.
Although, compared with (\ref{eq:RG}), the energy loss is too large, we find that the
elastic scattering cross sections $\psi_- e ({\rm He}) \to \psi_- e ({\rm He})$ are also large enough
to trap the DM inside red giants. The necessary cross section is about 1.6 pb~\cite{Davidson:1991}.
The cross section of   $\psi_- e\to \psi_- e$  is given by
\bea
 \sigma(\psi_- e\to \psi_- e) &=& \frac{f^2 m_e^2 \cos^2\alpha_H \sin^2\alpha_H}{32 \pi s v_H^2} \nonumber\\
&\times&\left[1+{4 (m_e^2-m_{\psi_-}^2) \over s \beta} \log {m_{\psi_-}^2 + s \beta \over m_{\psi_-}^2}\right] \nonumber\\
&\approx& 3 \times 10^{-4} \,{\rm pb} \ll 1.6 \,{\rm pb}
\eea
where $\beta=[s-(m_e+m_{\psi_-})^2][s-(m_e-m_{\psi_-})^2]/s^2$.
The electron contribution is too small.
However, the He contribution is more than $(m_{\rm He}/m_e)^4 \approx 10^{13}$ larger than the electron cross section,
even considering the order one uncertainty of scalar-He coupling, which is large enough  to keep the DMs from escaping red giants. We have checked
that similar arguments apply to other astrophysical objects.

\section{Decaying DM Scenario}
Another possibility in the Weinberg model is to utilize $\psi_+ \rightarrow \psi_{-} \alpha$ 
followed by $\alpha \rightarrow \gamma \gamma$ by some additional new physics 
acting on the Goldstone boson $\alpha$ that is described by effective Lagrangian:
\begin{equation}
{\cal L}_{\alpha \gamma \gamma} = \frac{c_{\alpha\gamma\gamma} \alpha_{\rm em}}{4 \pi  v_\varphi} 
\alpha F_{\mu\nu} \widetilde{F}^{\mu\nu}  
\end{equation} 
where $c_{\alpha\gamma\gamma}$ is a model dependent coupling which encodes underlying new 
physics for $\alpha \rightarrow \gamma\gamma$.
If two Majorana states $\psi_+$ and $\psi_{-}$ are quite heavy with small mass 
difference $\Delta M \sim 7.1$ keV, one can 
produce photon spectrum
from $\psi_+ \rightarrow \psi_{-} + \alpha \rightarrow \psi_{-} + \gamma \gamma$.
However we have to discuss first the explicit global symmetry breaking which would
generate nonzero finite mass for the Nambu-Goldstone boson $\alpha$.

\subsection{Explicit global symmetry breaking}

Note that $\alpha \rightarrow \gamma \gamma$ decay requires nonzero mass for 
the Goldstone boson $\alpha$. This seems contradictory to the model with 
spontaneously broken global dark symmetry, which would imply a massless Nambu-Goldstone (NG)
boson $\alpha$. However this is not really the case. Any global symmetry is expected
to be broken by gravity, so that global symmetry would be only approximately 
conserved. Small explicit breaking of spontaneously broken global symmetry 
would generate small but nonzero finite mass for the NG boson $\alpha$.
The explicit global symmetry breaking term can be organized as follows:
\begin{equation}
{\cal L}_{\rm sym- breaking} = - \frac{1}{2} m_\alpha^2 \alpha^2 
- \frac{\lambda_\alpha}{4 !} \alpha^4 + .....
\end{equation}

One can consider spurion technique to write down the explicit global 
symmetry breaking term, assuming a spurion $\chi$ with $U(1)_X$ charge $-1$ 
and mass dimension equal to zero so that $\chi \varphi$ is $U(1)_X$ invariant
and dim-1 operator. After constructing the $U(1)_X$ invariant Lagrangian,  
we set $\chi = v_\varphi / M_{\rm pl}$ and then global $U(1)_X$ will be 
explicitly broken due to quantum gravity effects. Note that $\chi \rightarrow 0$
as $M_{\rm pl} \rightarrow \infty$, the full global symmetry would be recovered
in this limit as anticipated. 

Then the leading operators with positive powers of $\chi$ that would break 
global symmetry explicitly due to quantum gravity effects would be 
\begin{widetext}
\begin{equation}
{\cal L}_{\rm spurion} = c_2 M_{\rm pl}^2 ( \chi \varphi )^2 + 
c_4 ( \chi \varphi )^4 + c_{2H} H^\dagger H ( \chi \varphi )^2 + H.c. ...
\end{equation}
\end{widetext}
where $c_i$'s are $\sim O(1)$ dimensionless couplings  
using the naive dimensional analysis. We assume that $c_i$'s are real.
Note that the decay $\phi \rightarrow \alpha \alpha$ is allowed by 
explicit global symmetry breaking, if it is kinematically allowed.

Under our assumption $\chi = v_\varphi / M_{\rm pl}$, only the 
$c_2$ term plays an important role, whose effects can be parametrized
in term of a single coupling $c_2$. And $c_2$ can be traded with the 
Goldstone boson mass $m_\alpha$ by the relation $m_\alpha^2 \equiv 6 c_2 v_\varphi^2 $. 
The NG boson mass $m_\alpha$, 
$\phi \rightarrow \alpha \alpha , 4 \alpha, 6 \alpha$, {\it etc}. 
and $\alpha^4$ self-interaction as well as $\phi \phi \rightarrow 
\alpha \alpha , 4 \alpha, $ {\it etc.} are all correlated with the coupling $c_2$. 

Once $\alpha$ gets massive through explicit global symmetry breaking, 
it would be able to decay into $\gamma\gamma$ and any other SM particles 
if kinematically allowed. Motivated by 3.5 keV X-ray line, we will consider
the decay chains, 
$\psi_+ \rightarrow \psi_{-} + \alpha \rightarrow \psi_{-} + \gamma \gamma$,
assuming $\Delta M \sim 7.1$ keV.

%

\subsection{Long lived $\psi_+$}
If $\psi_{+}$ could be cosmologically stable, we are led to consider 
two-component DM model with both $\psi_{-}$ and $\psi_+$. 
In order that it could explain the observed X-ray lines, the following 
condition is to be fulfilled:
\begin{equation}
\Gamma ( \psi_{+} \rightarrow \psi_{-} \gamma \gamma ) \simeq (3.3 \times 10^{-48} -5.9 \times 10^{-47})\,{\rm GeV}\times
\left( \frac{m_{\rm DM}}{\rm GeV} \right).
\label{eq:decayWidth}
\end{equation}
We note, however, that the above 3-body decay process gives broad photon spectrum peaked
around 3.55 keV rather than the observed line shape, for mass difference $\Delta M=7.1$ keV.
The decay rate of this mode is roughly order of 
\bea
\Gamma &\approx&
 \frac{c_{\alpha\gamma\gamma}^2 \alpha_{\rm em}^2 (\Delta M)^9}{192 \pi^5 m_\alpha^4 v_\varphi^4} \nonumber\\
&\approx& 4.2 \times 10^{-48} \,{\rm GeV} \nonumber\\
&\times& \left(c_{\alpha\gamma\gamma} \over 1\right)^2 
\left(\Delta M \over 7.1 \,{\rm keV}\right)^9
\left(10\,{\rm keV} \over m_\alpha\right)^4
\left(10^3\,{\rm GeV} \over v_\varphi\right)^4
\label{eq:3width}
\eea
assuming the $\alpha$ is more massive than the mass difference $\Delta M$. 
Both $\psi_+$ and $\psi_-$ can be thermalized by Higgs portal interaction and the emission of Goldstone boson,
$\psi_i \psi_i \rightarrow H_k H_l , \alpha \alpha$ ($i=+$ or$-$, and $k,l=1,2$), as well as the coannihilation channel $\psi_+ \psi_{-} \rightarrow H_{k=1,2} \alpha$. 
\begin{figure}
\center
\includegraphics[width=8cm]{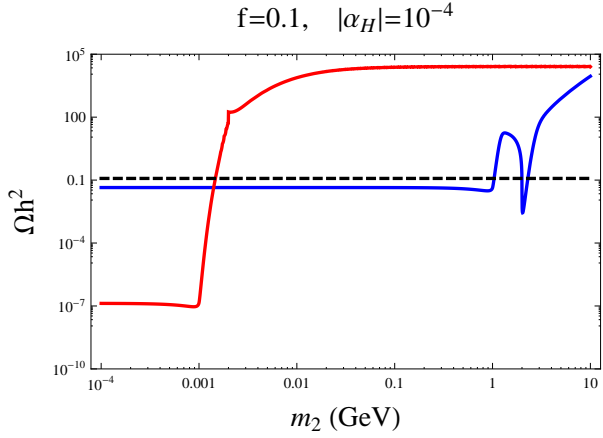}
\caption{The $\Omega_{\rm DM} h^2$ as a function of $m_2$ for $(m_{\rm DM},v_{\varphi})=(1 \,{\rm MeV}, 500\,{\rm GeV})$ (red line)
$(m_{\rm DM},v_{\varphi})=(1 \,{\rm GeV}, 3000\,{\rm GeV})$ (blue line). For other parameters, we assumed
$f=0.1, |\alpha_H|=10^{-4}$, $c_{\alpha\gamma\gamma}=1$, $\Delta M = 7.1$ keV, and $m_\alpha=10$ keV.}
\label{fig:om}
\end{figure}
Fig.~\ref{fig:om} shows the thermal relic density as a function of the $H_2$ mass $m_2$, assuming the
the mass difference $\Delta M = 7.1$ keV, the coupling $c_{\alpha\gamma\gamma}=1$,
the Goldstone boson mass $m_\alpha=10$ keV for two different values of DM mass $m_{\rm DM}=1$ MeV and 1 GeV.
For $m_{\rm DM}=1$ MeV (1 GeV) we take $v_\varphi \approx 3000 (500)$ GeV by comparing (\ref{eq:decayWidth}) and
(\ref{eq:3width}). We can see that the relic abundance can be easily accommodated.

The photons produced in the decay of the long lived $\psi_+$ inject energy to the baryon-photon plasma at the epoch of CMB decoupling.
Since such an energy injection disturb CMB photon and reionization history, it is highly constrained by the present CMB data \cite{Slatyer:2009yq,Cline:2013fm,Madhavacheril:2013cna}.
The constraint can be interpreted as a bound on the decay rate of the dark matter:
\bea \label{GammaCMBbnd}
\Gamma_+ 
&\lesssim& 10^{-10} \frac{\rho_{\rm DM}}{\GeV^3} \l( \frac{m_+}{E_\gamma} \r) \l( \frac{\rho_{\rm DM}}{\rho_+} \r) \l( \frac{T_{\rm cmb}}{T_0} \r)^3
\nonumber \\
&\simeq& 2 \times 10^{-43} \GeV \l( \frac{m_+}{1 \GeV} \r) \l( \frac{7 \keV}{E_\gamma} \r) \l( \frac{\rho_{\rm DM}}{\rho_+} \r)
\eea
where $\rho_{\rm DM}$ is the present energy density of dark matter, $E_\gamma$ is the injected energy from a decay of $\psi_+$,
$\rho_+$ is the present energy density of $\psi_+$, and we used $\Omega_{\rm DM} = 0.268$, 
$T_{\rm cmb} = 0.26 \eV$ and $T_0 = 2.35 \times 10^{-4} \eV$.
The decay width for X-ray signal, (\ref{eq:decayWidth}), satisfies the above bound for $m_{\rm DM} \lesssim 10^3$ GeV.

\section{Conclusions}
We proposed annihilating and decaying dark matter scenarios to explain the current dark matter relic abundance and
recently observed unidentified 3.55 keV X-ray line signal. In these scenarios we introduce  a global symmetry $U(1)_X$, and
also a Dirac fermion $\psi$, a scalar $\varphi$ which are charged under $U(1)_X$.
The scalar field $\varphi$ gets a vacuum expectation value $v_{\varphi}$ with a value in the range 10$^3$--10$^{16}$ GeV,
which breaks the $U(1)_X$ down to $Z_2$ symmetry under which $X \to -X$. 
The remnant $Z_2$ symmetry guarantees the stability of dark matter which we take to be $\psi_-$.
After the symmetry breaking
the Dirac fermion $\psi$ splits into two Majorana fermions $\psi_\pm$ with masses $m_{\psi_\pm}$.
The  real component of dark scalar $\varphi$, $\phi$, mixes with the SM Higgs $h$ with mixing angle 
$\alpha_H$ making two scalar
mass eigenstates $H_i (i=1,2)$ with masses $m_i$.

In the annihilating DM scenario, the monochromatic X-ray signal can be explained for the mixing angle $|\alpha_H| \sim 10^{-8} -10^{-2}$, 
the Yukawa coupling of the dark matter and the dark scalar
$f \sim 10^{-8}-10^{-3}$ when the resonance condition $m_2 = 2 m_{\psi_-}$ is satisfied with an accuracy at the
level of $10^{-3}$.
In the early universe where large relativistic degrees of freedom ($\sim 10^3$) which possibly comes from new 
physics contributions were present (entropy production processes during
the evolution of the universe can reduce this number), 
the process $\psi_- \psi_- \leftrightarrow  \alpha \alpha $ thermalizing $\psi_-$ with
the thermal plasma decouples around temperature $T_f =10^3-10^{10}$ GeV. Then the energy of dark matter simply
red shifts and becomes nonrelativistic at the temperature below its mass as the universe expands and cools down.
The Goldstone bosons 
may decouple from the SM thermal bath at temperature just above muon (electron) annihilation and contribute 
to the effective neutrino number as large as 0.39 (0.57).

Another possibility to give the X-ray signal in the model comes from the decay  of
long-lived $\psi_+$: $\psi_+ \to \psi_- \alpha \to \psi_- \gamma\gamma$, although the spectrum is not sharp.
We can obtain the required decay width for $\Delta M=m_{\psi_+}-m_{\psi_-} \approx 7.1$ keV
for the decaying DM mass $m_{\rm DM}\approx m_{\psi_\pm}$=1 GeV (1 MeV) when
$v_\varphi =500 (3000)$ GeV, $m_\alpha=10$ keV, $f=0.1$, $|\alpha_H|=10^{-4}$.


\begin{acknowledgments}
This work was supported by NRF Research Grant 
2012R1A2A1A01006053, and by SRC program of NRF 
Grant No. 20120001176 funded by MEST through 
Korea Neutrino Research Center at Seoul National University.
\end{acknowledgments}


\end{document}